\documentclass{article}

\usepackage{arxiv}

\usepackage[utf8]{inputenc} 
\usepackage[T1]{fontenc}    
\usepackage{hyperref}       
\usepackage{url}            
\usepackage{booktabs}       
\usepackage{amsfonts}       
\usepackage{nicefrac}       
\usepackage{microtype}      
\usepackage{lipsum}		
\usepackage{graphicx}
\usepackage{natbib}
\usepackage{doi}
\usepackage{amsmath} 
\usepackage{tabularx}

\title{Oral squamous cell detection using deep learning}

\author{ \href{https://orcid.org/0009-0009-7647-0731}{\includegraphics[scale=0.06]{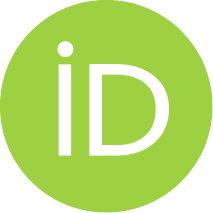}\hspace{1mm}Samrat Kumar Dev Sharma}\thanks{Use footnote for providing further
		information about author (webpage, alternative
		address)---\emph{not} for acknowledging funding agencies.} \\
	Department of Statistics\\
	Jagannath University\\
	9-10 Chittaranjan Ave, Dhaka 1100, Bangladesh \\
	\texttt{samrat.stat18@gmail.com} \\
}

\renewcommand{\shorttitle}{\textit{Oral squamous cell detection}}

\hypersetup{
pdftitle={Oral squamous cell detection using deep learning},
pdfsubject={q-bio.NC, q-bio.QM},
pdfauthor={Samrat Kumar Dev Sharma},
pdfkeywords={Deep Learning, Machine Learning, Oral squamous},
}

\begin{document}
\maketitle

\begin{abstract}
Oral squamous cell carcinoma (OSCC) represents a significant global health concern, with increasing incidence rates and challenges in early diagnosis and treatment planning. Early detection is crucial for improving patient outcomes and survival rates. Despite advances in understanding its molecular mechanisms, late diagnosis remains a substantial challenge in OSCC management. Precision medicine approaches are essential for personalized treatment, and deep machine learning techniques offer promising avenues for enhancing early detection and reducing cancer-related mortality and morbidity. 
Deep learning, a subset of machine learning, has shown remarkable progress in extracting and analyzing crucial information from medical imaging data. This article explores the technical foundations and algorithms of deep learning for OSCC, focusing on its applications in cancer detection, image classification, segmentation, and synthesis, as well as treatment planning. Automated image analysis powered by deep learning can provide valuable support to pathologists and clinicians in making informed decisions for cancer patients.
EfficientNetB3, an advanced convolutional neural network architecture, has emerged as a leading model for image classification tasks, including medical imaging. Its superior performance, characterized by high accuracy, precision, and recall, makes it particularly promising for OSCC detection and diagnosis. EfficientNetB3 achieved an accuracy of 0.9833, precision of 0.9782, and recall of 0.9782 in our analysis. By leveraging EfficientNetB3 and other deep learning technologies, clinicians can potentially improve the accuracy and efficiency of OSCC diagnosis, leading to more timely interventions and better patient outcomes. This article also discusses the role of deep learning in advancing precision medicine for OSCC and provides insights into prospects and challenges in leveraging this technology for enhanced cancer care.
\end{abstract}


\section{Introduction}
Oral squamous cell carcinoma (OSCC) is one of the most prevalent forms of cancer globally, with its incidence steadily rising in many populations like \cite{coletta2020grand}. This type of cancer affects the mouth and throat, posing significant challenges in terms of early detection and effective treatment planning. OSCC is particularly concerning due to its high incidence rate, late-stage diagnosis, and suboptimal treatment outcomes, all of which contribute to a high mortality rate. Effective management and early detection are crucial for improving prognosis, treatment efficacy, and survival rates for patients.
Oral squamous cell carcinoma (OSCC) accounts for 90\% of all oral cancers and represents a significant global health concern due to its high incidence and mortality rates \cite{jiang2019tobacco}. It can affect any anatomical site in the mouth, most commonly the tongue and the floor of the mouth. OSCC often arises from pre-existing potentially malignant lesions or de novo within a field of precancerous epithelium. Each year, approximately 405,000 new cases of oral cancer are reported worldwide, making it the seventh most common type of cancer in some regions, such as Brazil \cite{montero2015cancer}. 
The etiology of OSCC is multifactorial, with tobacco and alcohol consumption being the most significant risk factors, especially when used synergistically \cite{pinnika2024analysis}. Additional risk factors include poor diet (low in fruits and vegetables), betel nut chewing (prevalent in Asian populations), poor oral hygiene, and excessive exposure to ultraviolet light (for lip carcinomas). There is also evidence that high-risk HPV infections contribute to the development of OSCC, although the prevalence and impact of this factor remain under debate.
Epidemiological data show a higher prevalence of OSCC among males, likely due to greater exposure to risk factors. Typically, OSCC affects older individuals, with a mean age of diagnosis around 62 years. Clinically, it often presents as a painless ulcer on the border of the tongue or the floor of the mouth. Important prognostic factors include the size of the tumor at diagnosis, the presence of metastases in regional lymph nodes, and the depth of tumor invasion. Treatment generally involves surgery, radiation, and chemotherapy, either singularly or in combination
Despite the advancements in understanding the molecular mechanisms of OSCC, early detection remains a significant challenge. The current diagnostic methods are often invasive, time-consuming, and dependent on the clinician's expertise, which can lead to late-stage diagnoses. Early detection is crucial for improving prognosis, treatment efficacy, and survival rates.
The advent of deep learning, a subset of artificial intelligence (AI) and machine learning, offers promising avenues for enhancing the early detection and diagnosis of OSCC. Deep learning models, particularly convolutional neural networks (CNNs), have demonstrated remarkable progress in medical image analysis, enabling the extraction and analysis of crucial information from medical images.

\section{Releted Work}
\label{sec:headings}
In recent years, a plethora of research studies have explored the application of deep learning methodologies for early detection, diagnosis, and prognosis of oral squamous cell carcinoma (OSCC) as well as other types of cancers. In this section, we conduct a thorough examination of the existing literature on OSCC detection and the integration of deep learning techniques in medical imaging. Our aim is to gain a comprehensive understanding of the landscape surrounding OSCC diagnosis, highlighting both conventional methods and novel approaches driven by advancements in deep learning technology. \cite{kim2019deep} collected data from 255 patients treated at a surgical department between 2000 and 2017. They used deep learning with DeepSurv to predict survival outcomes and compared it to two other models. DeepSurv (.781) performed the best, with higher accuracy in predicting survival compared to the other models. In another study by  \cite{jeyaraj2019computer}, a deep convolutional neural network (CNN) model was proposed for OSCC detection. Remarkably, the model achieved the highest accuracy of 91.4\% with a sensitivity of 0.94 and specificity of 0.91 when tested on a dataset consisting of 100 images. In another study by  \cite{bur2019machine}, data from the National Cancer Database (NCDB) involving 782 patients were collected. Subsequently, using test data from 654 patients, they proposed a model based on the decision forest algorithm. Impressively, their model achieved an area under the curve (AUC) of 0.840, indicating its efficacy in predicting outcomes related to oral squamous cell carcinoma. Moreover,\cite{das2017computational} applied support vector machine (SVM) and random forest classifiers. Their proposed methodology efficiently detected mitotic cells from histopathological images of OSCC with 89\% precision, 95\% recall or sensitivity, 97.35\% specificity, 96.92\% accuracy, 96.45\% AUC, and a 92\% F-score measure. In a study conducted by \cite{rahman2022histopathologic} biopsy data of Oral Squamous Cell Carcinoma (OSCC) were collected from Kaggle. They proposed the AlexNet model, which achieved an impressive accuracy of 90.06\% and a loss rate of 9.08\% in predicting OSCC. In the study conducted by  \cite{welikala2020automated}, the ResNet-101 model was proposed. They achieved the highest F1 score of 87.07\% with this model. \cite{shavlokhova2021deep}, tissue samples from 20 patients were collected. They proposed the MobileNet model, which achieved a sensitivity of 0.47 and a specificity of 0.96, indicating its ability to effectively detect oral squamous cell carcinoma. In the study by \cite{Albalawi2024}, a histopathological imaging database for oral cancer analysis was utilized, comprising 1224 images from 230 patients. Their customized deep learning model demonstrated significant success, achieving an impressive 99\% accuracy when tested on the dataset. In a study conducted by  \cite{welikala2020fine}, various models including AlexNet, DenseNet-169, DenseNet-201, and ResNet-18 were employed. They achieved accuracies of 87.05\%, 95\%, 93.5\%, and 75.95\%, respectively, using these models.\cite{Warin2022} employed various models including DenseNet-169, ResNet-101, SqueezeNet, and Swin-S. The study proposed that the AUC of multiclass image classification for the best CNN model, DenseNet-169, was 1.00, and 0.98 respectively.  \cite{flugge2023detection} conducted a study where they randomly collected 1406 clinical photographs (CPs) from the Department of Oral and Maxillofacial Surgery, Charité - Universitätsmedizin Berlin. They utilized YOLOv5, ResNet-152, DenseNet-161, Inception-v4, and EfficientNet-b4 for their analysis. \cite{warin2022performance} retrospectively collected 600 oral photograph images, comprising 300 images of oral potentially malignant disorders (OPMDs) and 300 images of normal oral mucosa. They employed DenseNet-121 and ResNet-50 convolutional neural network (CNN)-based classification models. The image classification using DenseNet-121 achieved a precision of 91\%, a recall of 89.51\%, an F1 score of 95\%, a sensitivity of 100\%, a specificity of 90\%, and an AUC of the ROC curve of 95\%. On the other hand, the image classification using ResNet-50 achieved a precision of 92\%, a recall of 98\%, an F1 score of 95\%, a sensitivity of 98.39\%, and a specificity of 91.67\%.
In their study, \cite{das2023automatic} utilized an oral dataset collected from a repository \cite{rahman2020histopathological} containing images of normal oral cavities and oral squamous cell carcinoma (OSCC). They experimented with several deep learning models including VGG16, VGG19, AlexNet, ResNet50, ResNet101, MobileNet, and InceptionNet. The proposed model achieved an impressive cross-validation accuracy of 97.82\%, demonstrating its efficacy in automatically classifying oral cancer data. Specifically, the performance of the models varied, with AlexNet, ResNet50, ResNet101, MobileNet, and InceptionNet showing promising accuracy rates of 88\%, 91\%, 89\%, 93\%, and 92\%, respectively. In contrast, VGG16 and VGG19 exhibited lower performance with accuracies of 74\% and 71\%, respectively.
The existing literature underscores the potential of deep learning techniques in improving the classification of oral squamous cell carcinoma (OSCC) without the need for manual feature extraction. Despite this potential, there is a scarcity of research focusing on histopathological images of oral cancer cells using deep learning approaches. The literature emphasizes the significance of deep learning methods for classification tasks, indicating a promising avenue for further exploration. Building upon this literature analysis, our proposed study aims to address the gap by focusing on the binary classification of oral histopathological images using various CNN architectures with different configurations and layers. By comparing the outcomes of these variations, we seek to identify the best-performing model for accurate classification.
In summary, the collective findings from these studies underscore the promising role of deep learning techniques in improving OSCC detection, diagnosis, and prognosis. This not only holds potential for enhancing patient outcomes but also contributes to the advancement of cancer research and treatment strategies.


\section{Methodology}
\label{sec:methodology}

\subsection{System Design}
\label{sec:system-design}

\begin{figure}[h!]
    \centering
    \includegraphics[width=\linewidth]{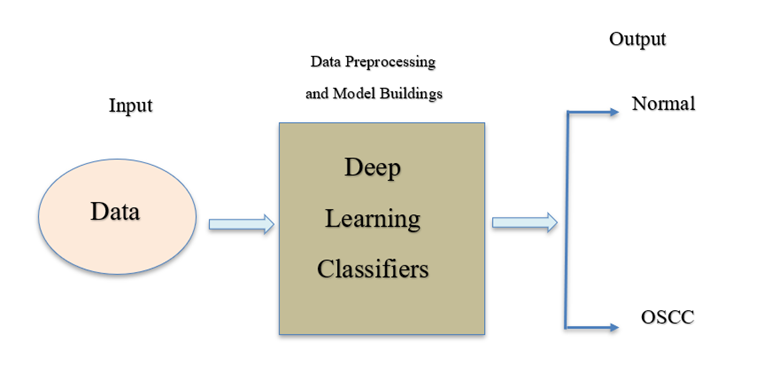} 
    \caption{System Design for the research.}
    \label{fig:fig1}
\end{figure}

\subsection{Workflow}
\label{sec:work-flow}
\begin{figure}[ht!]
    \centering
    \includegraphics[width=0.8\linewidth]{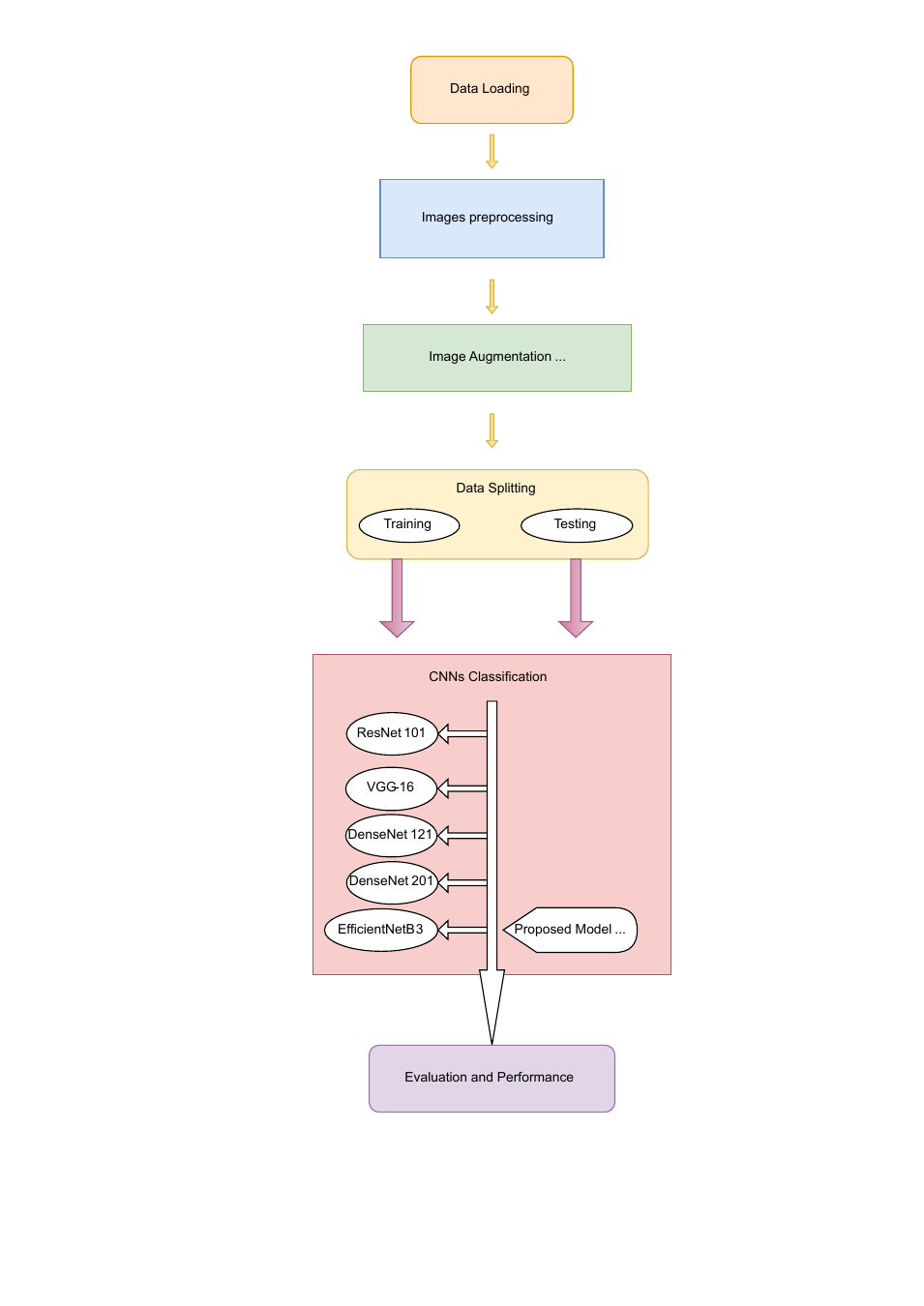} 
    \caption{Workflow}
    \label{fig:workflow}
\end{figure}

 According to the workflow illustrated in Figure \ref{fig:workflow}, the first step of this project involves data collection. Data for this project was sourced primarily from the Oral Cancer Database (OCDB) available on Kaggle,\cite{ashenafi2023} supplemented by additional high-resolution oral cell images obtained through collaboration with various hospitals. The dataset consists of images distributed across training, testing, and validation sets, each containing two categories: Normal and OSCC (Oral Squamous Cell Carcinoma). The training data comprises 2435 images labeled as Normal, and 2511 images labeled as OSCC. The testing data includes 31 images labeled as Normal and 95 images labeled as OSCC, while the validation data contains 28 images labeled as Normal and 92 images labeled as OSCC. In total, the dataset comprises 5192 images, with a distribution of approximately 52\% OSCC and 48\% Normal images.

\subsection{Image Processing}
\label{sec:Image-Processing}
\begin{figure}[h!]
    \centering
    \includegraphics[width=\linewidth]{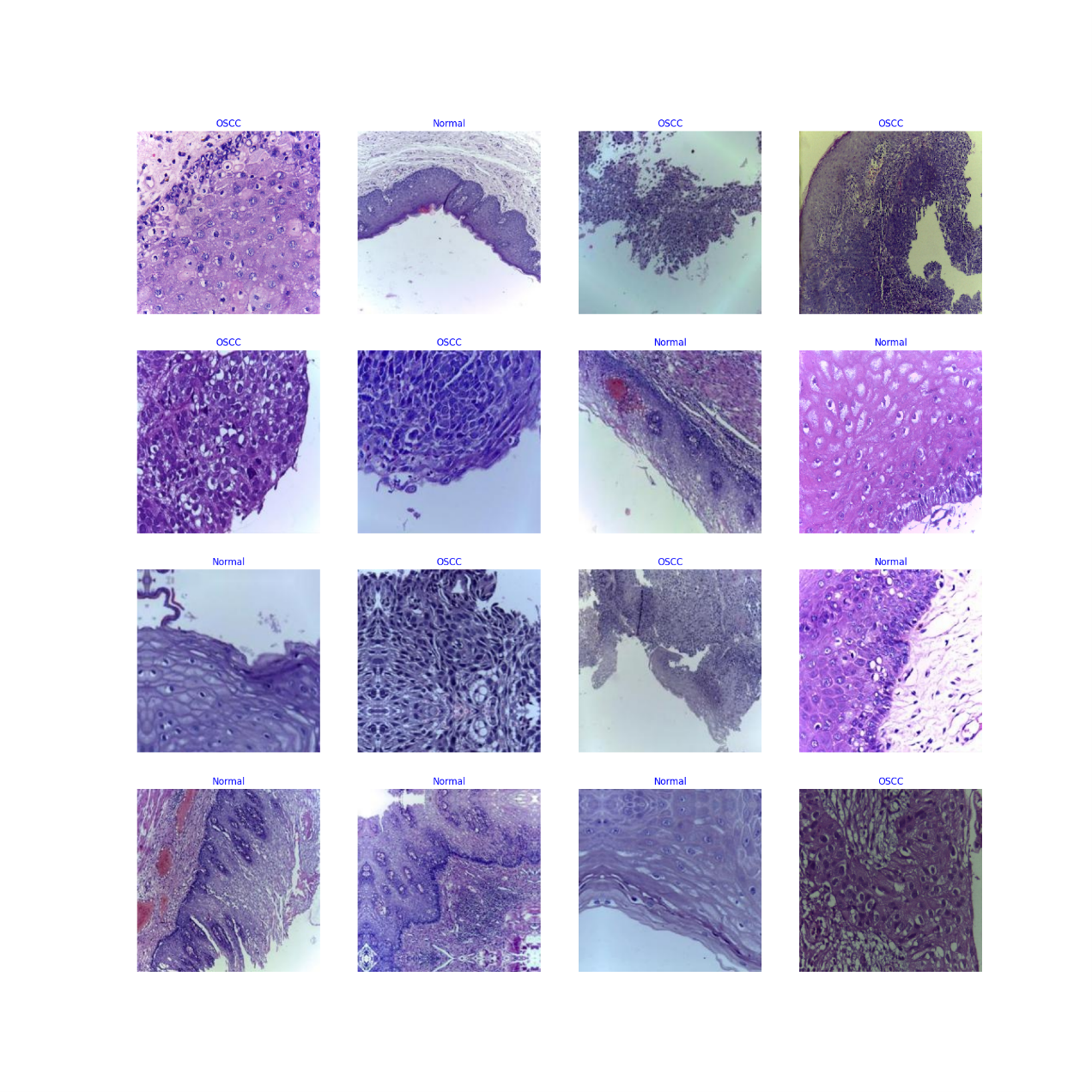} 
    \caption{Preprocessed Images}
    \label{fig:preprocessedimage}
\end{figure}

The collected images exhibit variability in terms of sizes, zoom levels, angles, lighting conditions, and orientations. To standardize the dataset for model training, all images underwent preprocessing steps. Specifically, they were manually cropped, resized to dimensions of 224 × 224 × 3, and converted into the JPG file format. This preprocessing ensures consistency in the input data format across all images, facilitating compatibility with the chosen deep learning model architecture. After the preprocessing of the images, the images are used for the image augmentation process like, zooming, right shifting, flipping etc. Figure \ref{fig:preprocessedimage}, illustrates the preprocessed data.

\subsection{Image Augmentation}
\label{sec:image-augmentation}
In this project, image augmentation is employed to enhance the diversity and robustness of the training dataset. Image augmentation is a technique that generates modified versions of images in the dataset, artificially expanding the size and variability of the dataset. This helps improve the model's performance and generalization ability by introducing variations that the model might encounter in real-world scenarios. Specifically, we apply horizontal flipping to the training images, which randomly flips images along the horizontal axis. This augmentation technique is critical for training deep learning models as it allows them to learn to recognize objects in images regardless of their orientation, thereby improving robustness. Additionally, all images were resized to a uniform size of 224x224 pixels and color channels were standardized to RGB format to maintain consistency across the dataset. For validation and testing datasets, only basic preprocessing was applied without any augmentation to ensure that the evaluation metrics reflect the model's performance on unaltered images. This augmentation strategy significantly contributes to the model's ability to generalize well to new, unseen data, which is crucial for achieving high accuracy in image classification tasks.
\subsection{Data set Spliting}
\label{sec:dataspliting}
The dataset used in this study was split into training, validation, and test sets to facilitate model development and evaluation. Initially, the dataset was divided into training and temporary test sets in a 70-30 ratio, ensuring that the class distribution was preserved across the splits. Subsequently, the temporary test set was further divided into validation and final test sets in a 50-50 ratio, maintaining the same class distribution. This strategy ensured that the training set was used to train the model, the validation set was employed for hyperparameter tuning and model selection, and the final test set served for unbiased evaluation of the trained model's performance. Each set was encapsulated in DataFrame objects, containing image paths and corresponding labels. This rigorous splitting approach ensured that the model's performance was robustly assessed on unseen data, contributing to the reliability and generalization ability of the developed model.
\subsection{Pre-trained DL-CNN Model}
\label{sec:pretrained Dl and CNN}
Deep Learning Convolutional Neural Network (DL-CNN) models have revolutionized various image-based tasks, including object detection and recognition. CNNs consist of layers interconnected in a sequential manner, where each layer is linked to the subsequent layer [6]. These layers consist of neurons, forming a spatial architecture that creates a volume with width, height, and depth. The depth of the network corresponds to the number of stacked layers, with each layer contributing to feature extraction and representation. Key components of a CNN architecture include convolutional, pooling, and fully connected layers, along with activation functions like ReLU, batch normalization, and dropout layers.
In a typical CNN, convolution layers comprise various filters with width, height, and depth, which extract different features from the input image. Each filter is parameterized and convolved over the input image to extract features via dot product operations. Parameters such as size, stride, and padding control the behavior of convolutional layers. Pooling layers further process the extracted features, reducing the size of the feature map to reduce computational complexity. Common pooling techniques include average pooling and max pooling. Batch normalization layers, along with activation functions like ReLU, help normalize shifts in middle layers, aiding in network convergence. Dropout layers mitigate overfitting by randomly dropping neurons during training. Finally, the reduced feature map is passed through fully connected layers with the SoftMax function for classification into corresponding classes.
Pre-trained DL-CNN models such as VGG-16 \cite{szegedy2016rethinking}, ResNet50 \cite{zoph2018learning}, Inception-V3, NASNetLarge , Xception \cite{chollet2017xception}, and DenseNet201 \cite{huang2017densely} are widely used for image classification tasks. These models, trained on large-scale datasets like ImageNet, demonstrate strong generalization capabilities when applied to external datasets. In this context ResNet101, VGG-16, DenseNet121, DenseNet201 and EfficientNetB3 are considered as candidates for pre-trained models, which are further modified with additional layers for effective Oral Squamous Cell Carcinoma (OSCC) detection. 
\subsubsection{ResNet}
\label{sec:resnet}
ResNet101 is a deep convolutional neural network architecture known for its depth and effectiveness in image recognition tasks. With 101 layers, it surpasses its predecessors, addressing the challenge of vanishing gradients through skip connections. These connections allow gradients to flow directly, enabling the training of very deep networks with improved convergence and accuracy. Each ResNet101 block consists of residual units, featuring multiple convolutional layers with batch normalization and ReLU activation. Additionally, it adopts a bottleneck design in deeper layers to reduce complexity while maintaining performance. Pre-trained ResNet101 models, often trained on datasets like ImageNet, excel in various computer vision tasks. Through its depth, skip connections, and innovative design, ResNet101 has significantly advanced deep learning in computer vision.

\subsubsection{VGG-16}
\label{sec:vgg16}
VGG-16 is a convolutional neural network created by the Visual Geometry Group (VGG) at the University of Oxford, known for its straightforward design and effectiveness in image classification. With 16 weight layers, including 13 convolutional layers and 3 fully connected layers, VGG-16 maintains a consistent structure. It processes 224x224 RGB images through convolutional blocks followed by max pooling layers to extract features and reduce dimensions. Each block contains multiple 3x3 convolutional layers with ReLU activation functions, increasing filter count for complex patterns. Fully connected layers with 4096 units and ReLU activation, along with dropout regularization, prevent overfitting. The final layer uses softmax activation to produce class probabilities. Despite its simplicity, VGG-16 performs well on image classification benchmarks, making it widely used in computer vision research.

\subsubsection{DenseNet121}
\label{sec:densenet121}
DenseNet121 is a unique convolutional neural network with 121 layers, part of the DenseNet family. Its standout feature is the dense connectivity between layers, where each one is directly linked to every other in a feed-forward manner. These fosters feature reuse, propagation, and gradient flow, aiding in better performance and convergence. Each dense block in DenseNet121 includes multiple convolutional layers, batch normalization, and ReLU activations for feature extraction. Transition layers between dense blocks reduce spatial dimensions and parameters. DenseNet121 effectively tackles the vanishing gradient problem and excels in various computer vision tasks like image classification and object detection. Pretrained on ImageNet, it's widely used in both research and practical applications for its compact design and efficient parameter utilization.

\subsubsection{DenseNet201}
\label{sec:DenseNet201}
DenseNet201, an extension of the DenseNet architecture, is a deep convolutional neural network known for its dense connectivity patterns and impressive performance in image recognition. With 201 layers, DenseNet201 features densely connected blocks where each layer receives input from all preceding layers, promoting feature reuse and gradient flow. These blocks contain multiple convolutional layers, batch normalization, and ReLU activations for effective feature extraction. Transition layers manage spatial dimensions and model complexity. Pretrained on datasets like ImageNet, DenseNet201 excels in tasks like image classification and object detection. Its depth, dense connectivity, and efficient parameter use make it a popular choice in both research and practical applications in deep learning and computer vision.

\subsubsection{EfficientNetB3}
\label{sec:efficientNetB3}
EfficientNetB3, part of the EfficientNet family, is a convolutional neural network celebrated for its efficiency and effectiveness in image recognition. Developed by Google AI researchers, it strikes a balance between model size, computational resources, and performance, making it ideal for resource-limited environments. Through compound scaling, EfficientNetB3 optimizes its depth, width, and resolution for superior performance. It features multiple blocks with inverted bottleneck structures, utilizing depthwise separable convolutions and squeeze-and-excitation modules to enhance feature representation efficiently. Pretrained on datasets like ImageNet, EfficientNetB3 excels in tasks such as image classification and object detection. Its compact design and ability to achieve high accuracy with fewer parameters make it a compelling option for both research and practical applications in deep learning and computer vision.
\subsection{Model Training}
\label{sec:model training}
All models were trained using the NVIDIA GeForce MX110 GPU, leveraging the Scikit-learn 1.4.2 library for machine learning tasks. This setup ensured efficient processing and utilization of computational resources during model training and evaluation.

\section{Experiment, Result, Analysis and Discussion}
\label{sec:result}
\subsection{Evaluation Measures}
\label{sec:evaluation result}
In our experimental study, we opted for five recently developed pre-trained deep learning convolutional neural network (DL-CNN) models—ResNet101, VGG-16, DenseNet121, DenseNet201, and EfficientNetB3—tailored for detecting oral squamous cell carcinoma (OSCC) in histopathological images of oral lesion biopsies.
Evaluation metrics are used to assess the performance of the DL-CNN models. It provides quantitative measurements that help in comparing different models and selecting the most suitable one for a particular task. Common evaluation metrics are included as part of the analysis, including accuracy, precision, recall, F1-score, and area under the receiver operating characteristic curve (ROC AUC) which is defined by the equations (1)-(4). Accuracy measures the proportion of correctly classified instances out of the total number of instances, providing an overall assessment of model performance. Precision quantifies the proportion of true positive predictions among all positive predictions, highlighting the model's ability to avoid false positives. Recall, also known as sensitivity, calculates the proportion of true positive predictions among all actual positive instances, indicating the model's ability to capture all relevant instances. F1-score, the harmonic means of precision and recall, balances both metrics, providing a comprehensive measure of model accuracy. Lastly, ROC AUC evaluates the model's ability to distinguish between positive and negative classes across various threshold values, with a higher AUC indicating better model performance. By considering these evaluation metrics collectively, researchers can make informed decisions regarding model selection and refinement to optimize performance for the specific task at hand.
\begin{itemize}
\item \textbf{Accuracy}: It measures the proportion of correctly classified instances out of the total number of instances. It can be formulated as:
\begin{equation}
\text{Accuracy} = \frac{TP + TN}{TP + TN + FP + FN}
\end{equation}
Where:

\begin{itemize}
    \item $\text{TP:}$ is the number of true positives (correctly predicted positive instances).
    \item $\text{TN:}$ is the number of true negatives (correctly predicted negative instances).
    \item $\text{FP:}$ is the number of false positives (incorrectly predicted as positive instances).
    \item $\text{FN:}$ is the number of false negatives (incorrectly predicted as negative instances).
\end{itemize}
\item \textbf{Precision}: It measures the proportion of correctly predicted positive instances out of all instances predicted as positive. It can be formulated as:
\begin{equation}
\text{Precision:} = \frac{TP}{TP + FP}
\end{equation}
\item \textbf{Recall}: It measures the proportion of correctly predicted positive instances out of all actual positive instances. It can be formulated as:
    \begin{equation}
        \text{Recall} = \frac{TP}{TP + FN}
    \end{equation}
    
    \item \textbf{F1-Score}: It is the harmonic mean of precision and recall, providing a balance between the two metrics. It can be formulated as:
    \begin{equation}
        \text{F1-Score} = \frac{2 \times \text{Precision} \times \text{Recall}}{\text{Precision} + \text{Recall}}
    \end{equation}
 
\end{itemize}

\subsection{Comparison of various DL-CNN models results}
\label{sec:results-comparisons}
\begin{figure}[ht!]
    \centering
    \includegraphics[width=\linewidth]{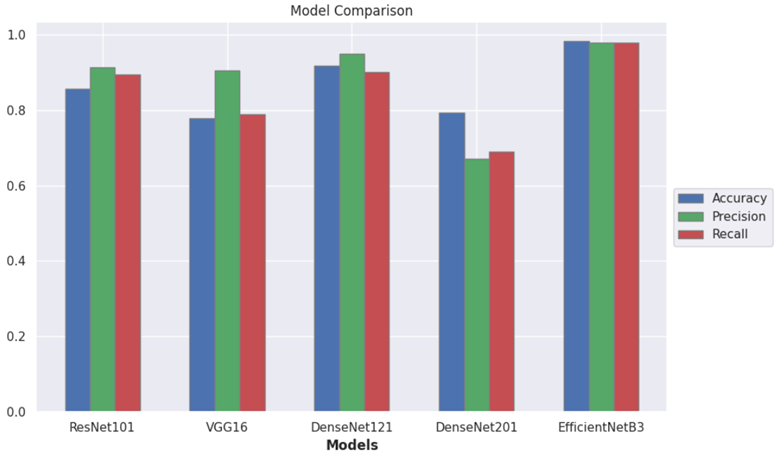} 
    \caption{Performance Different Deep Learning CNNs Model}
    \label{fig:performace}
\end{figure}

In our comparative analysis, we evaluated the performance of five candidate pre-trained DL-CNN models—ResNet101, VGG-16, DenseNet121, DenseNet201, and EfficientNetB3—each modified with additional layers, on the selected datasets. The evaluation encompassed various metrics including accuracy, precision, recall, and loss, shedding light on their classification capabilities. In Figure \ref{fig:performace}, we present the performance of different deep learning CNN models based on the evaluation metrics.

ResNet101 demonstrated competitive performance with an accuracy of 0.8571, precision of 0.9140, recall of 0.8947, and a loss of 0.4435. VGG-16 exhibited moderate accuracy (0.7778) and recall (0.7895), with higher precision (0.9036) and a slightly lower loss of 0.4415 compared to ResNet101. DenseNet121 showcased superior accuracy (0.9178) and precision (0.9481), along with a commendable recall of 0.9012 and a loss of 0.6411. Conversely, DenseNet201 yielded lower performance metrics, with an accuracy of 0.7936, precision of 0.6700, recall of 0.6900, and a significantly higher loss of 2.2123.

EfficientNetB3 emerged as the top-performing model, surpassing its counterparts with an outstanding accuracy of 0.9833, precision of 0.9782, recall of 0.9782, and the lowest loss of 0.1117. Its exceptional performance can be attributed to the densely connected CNN layers, which optimize data flow and mitigate gradient-related issues. Furthermore, EfficientNetB3's parameter efficiency and regularization effects contribute to reduced overfitting, making it highly effective even with limited training data. Table 1 shows the models comparisons

\begin{table}[ht]
  \centering
  \caption{Results of Oral Cancer Database (OCDB)}
  \label{tab:results} 
  \begin{tabular}{lcccc}
    \hline
    \textbf{Models} & \textbf{Loss} & \textbf{Accuracy} & \textbf{Precision} & \textbf{Recall} \\
    \hline
    ResNet101 & 0.4435 & 0.8571 & 0.9410 & 0.8947  \\
    VGG-16 & 0.4415 & 0.7778 & 0.9036 & 0.7895 \\
    DenseNet121& 0.6411 & 0.9178 & 0.9481 & 0.9012  \\
    DenseNet201 & 2.2123 & 0.7936 & 0.6700 & 0.6900 \\
    EfficientNetB3 & 0.1117 & 0.9833 & 0.9782 & 0.9782 \\
    \hline
  \end{tabular}
\end{table}

\subsection{Result Analysis Using Performance Measuring Graph}
\label{sec:results analysis}
\begin{figure}[ht!]
    \centering
    \includegraphics[width=\linewidth]{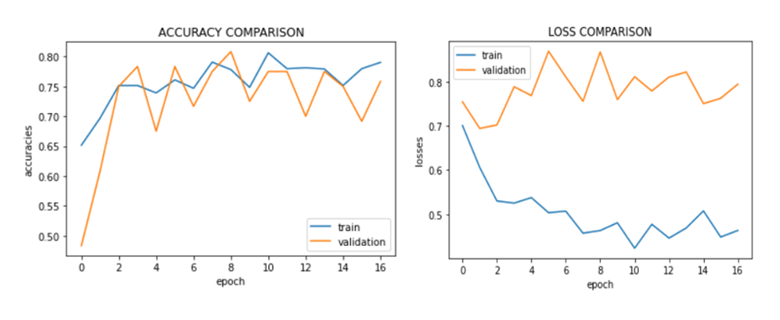} 
    \caption{Accuracy and Loss of model VGG-16}
    \label{fig:vgg-16}
\end{figure}

In this section, the results analysis of the proposed model is doing using a performance measuring graph, especially the accuracy and loss graph, generated during the training and validation process for all the comparative models. A model with minimum loss signifies the best result. The minimum loss indicates that the model learns from the training and validation phase with a lower error rate. On the other hand, the maximum accuracy value indicates optimal results for the model.
In Figure \ref{fig:vgg-16}, the VGG-16 model's accuracy graphs reveal similar training and validation accuracies, suggesting no overfitting briefly. However, the loss graphs present a different perspective. When the training loss is 0.46, the training accuracy is 0.79. For the validation data, the loss is 0.51, and the accuracy is 0.75. This disparity indicates that while the model's accuracy is good, the higher validation loss compared to training loss suggests some degree of overfitting. The model performs well on the training data but slightly less so on the validation data.

\begin{figure}[ht!]
    \centering
    \includegraphics[width=\linewidth]{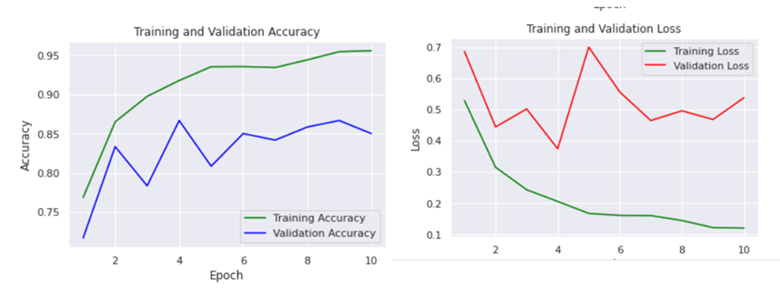} 
    \caption{Accuracy and Loss of model ResNet101}
    \label{fig:resnet101}
\end{figure}

\begin{figure}[ht!]
    \centering
    \includegraphics[width=\linewidth]{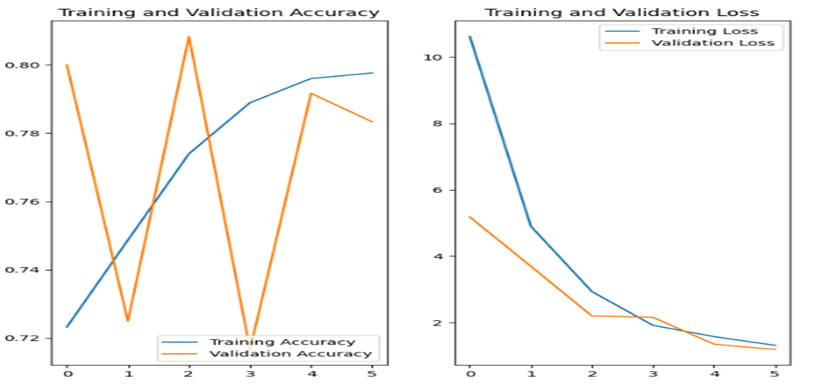} 
    \caption{Accuracy and Loss of model DenseNet201}
    \label{fig:densenet201}
\end{figure}

In Figure \ref{fig:resnet101}, ResNet101 model trained on the ImageNet dataset shows high performance with a training accuracy of 96\% and a validation accuracy of 93\%. The training loss decreased from 0.7 to 0.2, and the validation loss decreased from 0.6 to 0.3. Despite the high accuracy, the lower training loss compared to validation loss suggests that the model is overfitting to the training data.
In Figure \ref{fig:densenet201}, the DenseNet201 model achieved a training accuracy of about 80\% and a validation accuracy of about 79\%. The training loss decreased significantly from about 10 to 2, while the validation loss decreased from about 10 to 5. These results indicate that the model is learning effectively and not overfitting, maintaining a good balance between training and validation performance.

\begin{figure}[ht!]
    \centering
    \includegraphics[width=\linewidth]{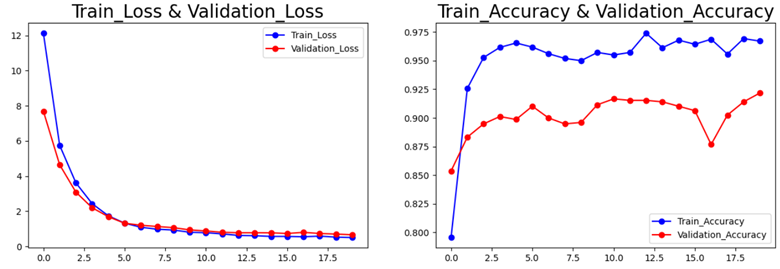} 
    \caption{Accuracy and Loss of model DenseNet121}
    \label{fig:densenet121}
\end{figure}

In Figure \ref{fig:densenet121}, the DenseNet121 model was trained for 20 epochs. The training loss decreased dramatically from 11.5 to 0.5, while the validation loss decreased from 5.5 to 1.2. The training accuracy increased from 0.2 to 0.95, and the validation accuracy increased from 0.8 to 0.9. The model achieved an excellent performance on the validation set, with an accuracy of 90\%, indicating that it can learn the features of the images effectively and classify them correctly without overfitting.

\begin{figure}[ht!]
    \centering
    \includegraphics[width=\linewidth]{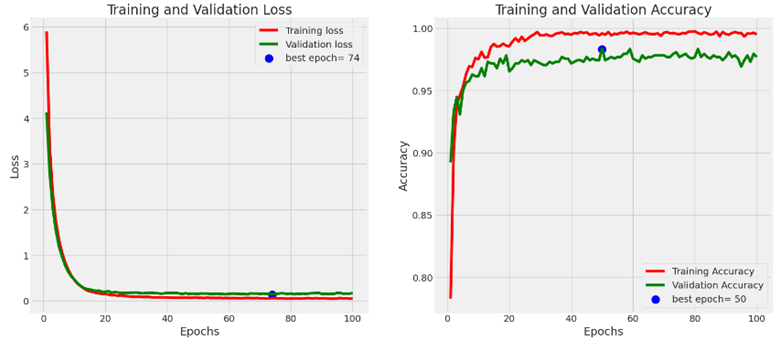} 
    \caption{Accuracy and Loss of model EfficientNetB3}
    \label{fig:efficient}
\end{figure}

In Figure \ref{fig:efficient}, the EfficientNetB3 model achieved its best validation loss of 0.05 at epoch 74 and its best validation accuracy of 0.99 at epoch 50. The training loss and accuracy curves indicate that the model learns the training data well and does not overfit, maintaining high performance across both training and validation sets.

Overall, the models exhibit varying degrees of overfitting and learning efficiency. The VGG-16 and ResNet101 models show some signs of overfitting, with higher validation losses compared to training losses. In contrast, the DenseNet201 and DenseNet121 models demonstrate effective learning with minimal overfitting, maintaining good performance on both training and validation data. The EfficientNetB3 model stands out with exceptional performance and no signs of overfitting, achieving near-perfect validation accuracy and very low validation loss. These observations highlight the importance of analyzing both accuracy and loss metrics to fully understand model performance and generalization capabilities.

\section{Future work}
\label{sec:future work}
To build upon our findings and further enhance the automated detection and diagnosis of oral squamous cell carcinoma (OSCC), several avenues for future improvement can be considered. Firstly, expanding the datasets used for training and validation to encompass larger and more diverse populations could enhance the robustness and generalizability of the models. Additionally, implementing more advanced data augmentation techniques can help in creating more diverse training samples, potentially improving model performance. Furthermore, conducting further hyperparameter optimization, including fine-tuning learning rates, batch sizes, and the number of training epochs, could lead to better model performance. Exploring ensemble learning techniques by combining multiple models through ensemble methods may leverage the strengths of individual models, resulting in improved detection accuracy. Moreover, applying transfer learning from models trained on large, diverse datasets could enhance the detection capabilities of histopathological images, especially for rare or complex cancer types. Extending the model to detect distinct stages of oral cancer could provide valuable insights for both patients and doctors, aiding in more effective treatment planning and prognosis assessment. Finally, conducting extensive clinical trials to validate the models' performance in real-world settings is crucial for their adoption in medical practice. By addressing these limitations and exploring the suggested improvements, future research endeavors can advance the field of automated cancer detection, ultimately leading to better patient outcomes and more efficient treatment processes.

\bibliographystyle{unsrtnat}
\bibliography{references}  






\end{document}